\documentclass[preprint,12pt]{elsarticle} 
\usepackage{graphicx} 
\usepackage{amssymb} 
\usepackage{amsmath} 

\journal{Nuclear Physics A} 

\begin{document} 

\begin{frontmatter} 

\title{Nuclear Structure and the Fate of \\ Core Collapse (Type II) Supernova} 
\author{Moshe Gai} 
\address{LNS at Avery Point,University of Connecticut~,~Groton,~CT 06340-6097, USA \\
and Wright Lab, Dept of Physics, Yale University, New Haven, CT 06520-8124, USA} 

\begin{abstract} 
For a long time Gerry Brown and his collaborator Hans Bethe considered the question of the final fate of a core collapse (Type II) supernova. Recalling ideas from nuclear structure on Kaon condensate and a soft equation of state of the dense nuclear matter they concluded that progenitor stars with mass as low a 17-18M$_\odot$ (including supernova 1987A) could collapse to a small mass black hole with a mass just beyond 1.5M$_\odot$, the upper bound they derive for a neutron star. We discuss another nuclear structure effect that determines the carbon to oxygen ratio (C/O) at the end of helium burning. This ratio also determines the fate of a Type II supernova with a carbon rich progenitor star producing a neutron star and oxygen rich collapsing to a black hole. While the C/O ratio is one of the most important nuclear input to stellar evolution it is still not known with sufficient accuracy. We discuss future efforts to measure with gamma-beam and TPC detector  the $^{12}C(\alpha,\gamma)^{16}$O reaction that determines the C/O ratio in stellar helium burning.

\end{abstract}


\end{frontmatter} 

\section{Dedication} 
\label{sec:ded} 

This work is dedicated to the memory of my beloved teacher and friend Gerry Brown who taught me how to think and approach a problem. I consider myself very lucky to {\bf have been under the wings of this man who flew with the eagles} \cite{Eagles}.
\newpage 

\section{Introduction}

By the year 1985 Gerry Brown and his collaborator Hans Bethe were able to compile three decades of work into a self containing comprehensive picture of core collapse (Type II) supernova explosions that could be even conveyed to a general audience in the popular publication of the Scientific American \cite{SciAm}. The onion layer structure of a 25 solar mass progenitor star just before it collapses under its own gravity and the temperature, density and duration of each prior burning stage are shown in Fig. 1 which is taken from this Scientific American article \cite{SciAm}.

\begin{figure}[hbt] 
\begin{center} 
\includegraphics[width=\textwidth]{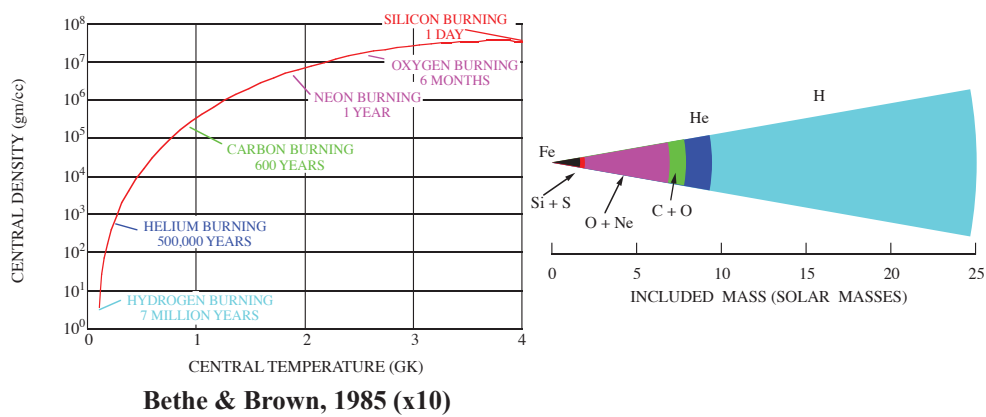} 
\caption{The onion layered structure of a 25 solar masses progenitor star before a core collapse (Type II) supernova and the time, temperature and density range of the various burning stages as illustrated by Bethe and Brown \cite{SciAm} with the temperature scale corrected by a factor of approximately 10 \cite{Gai1}.} 
\label{fig:TypeII}
\end{center} 
\end{figure} 

One of the central issue in the stellar evolution illustrated in Fig. 1 is whether Type II supernova end up as a black hole or a neutron star. In particular one would like to determine the smallest mass above which Type II supernova ends up as a black hole. Early on Gerry Brown considered the effect of kaon condensate on the maximum possible mass for a neutron star that he derived to be no more than 1.5M$_\odot$ \cite{Br88}. Kaon condensate he claimed leads to a soft equation of state \cite{Br92} of the dense nuclear matter and it gave further credence to his limit on the maximum mass of a neutron star. This lead Bethe and Brown \cite{Br94} to conclude that low mass black hole with mass just above 1.5M$_\odot$ may be very common and progenitors stars with mass as low as 17-18M$_\odot$, including supernova 1987A, ended up as a black hole. Indeed the minimum mass of a progenitor star that collapses to a black hole seems to be around 20M$_\odot$, close to the mass of the progenitor star of the supernova 1987A and as we discuss below other nuclear structure effect play a role in concluding the minimum mass of the progenitor star.

\section{Helium Burning and the C/O Ratio}

We consider here another nuclear structure effect that determines the fate of a Type II supernova vis-a-vis the C/O ratio at the end of helium burning. Stellar helium burning, the second burning stage that follows hydrogen burning, as shown in Fig. 1, is an important stage in the evolution of stars. During this stage the elements carbon and oxygen are formed and as such it is one of the most vivid examples of the anthropic principle \cite{Fowler}. During this stage carbon is synthesized by the ``triple-$\alpha$ process" via the Hoyle state at 7.654 MeV in $^{12}$C but at the same time carbon is also destroyed by fusing with an additional alpha-particle to form $^{16}$O in the $^{12}$C$(\alpha,\gamma)^{16}$O reaction. Hence the formation of oxygen in stellar helium burning determines the C/O ratio; an essential parameter in stellar evolution theory \cite{Fowler}. 

Stellar evolution theory requires the knowledge of the C/O ratio with an uncertainty of 5\%. This requires accurate measurements at low energies and extrapolation of the measured astrophysical cross section factors to the Gamow window at 300 keV \cite{Fowler}. Since mainly two ($\ell$ = 1 and $\ell$ = 2) partial waves contribute to the reaction, accurate angular distribution data are needed at low energies to determine with high accuracy the astrophysical cross section factors $S_{E1}(300)$ and $S_{E2}(300)$ defined in Ref. \cite{Fowler}.

The importance of the C/O ratio for the evolution of massive stars ($M  >  8M_\odot$) that evolve to core collapse (type II) supernova has been discussed extensively \cite{Weaver}. More recently it was shown that the C/O ratio is also important for understanding the $^{56}$Ni mass fraction produced by lower mass stars ($M \ \approx \ 1.4M_\odot$) that evolve into Type Ia supernova (SNeIa) \cite{SNeIa}. Thus the C/O ratio is also important for understanding the light curve of SNeIa. Such light curves of SNeIa are used as cosmological ``standard candles" with which the accelerated expansion of the universe and dark energy were recently discovered \cite{RMP}. \

\subsection{The $^{12}C(\alpha,\gamma)^{16}$O Reaction}

Recently some of the most impressive gamma-ray measurements of the $^{12}$C$(\alpha,\gamma)^{16}$O reaction were published \cite{Kunz,Ham1,Ham,Ham2,Plag} including measurements of complete angular distribution at center-of-mass energies approaching 1.0 MeV. These measurements employ large luminosities of the order of $10^{35}$  cm$^{-2}$sec$^{-1}$ with integrated luminosities close to one inverse fb \cite{Ham1,Ham,Ham2,Plag}, and a large (fraction of $4\pi$) array of gamma-ray detectors (but some of the arrays employ low-efficiency HpGe detectors, which led in some cases to insufficient counting statistics). Such unprecedented data with unprecedented characteristics led to an expectation of a resolution of the debate on the value of the low-energy cross section of the $^{12}$C$(\alpha,\gamma)^{16}$O reaction. While these data did not resolve the outstanding questions they provide the first possible detailed study of the cross section of the $^{12}$C$(\alpha,\gamma)^{16}$O reaction at low energies approaching 1.0 MeV.

\begin{figure}[hbt] 
\begin{center} 
\includegraphics[width=5in]{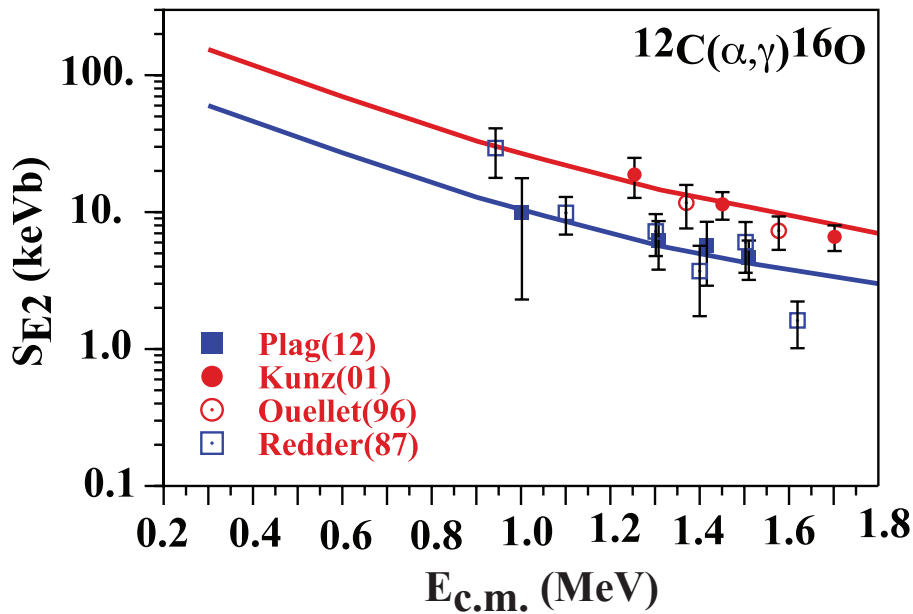} 
\caption{The measured $S_{E2}$ values \cite{Kunz,Plag,Redder,Ouellet} and the corresponding R-matrix fits. The two distinct groups of data extrapolate to $60 \pm 12$ and $154 \pm 31$ keVb. The $S_{E2}$ values measured using the GANDI \cite{Ham1} and EUROGAM \cite{Ham2} arrays are excluded, as discussed in Ref. \cite{Gai13}.}
\label{fig:E2}
\end{center} 
\end{figure} 

In Fig. 2 we show the published ``world data" \cite{Kunz,Plag,Redder,Ouellet} of $S_{E2}$ values deduced from angular distributions measured at low energies ($E_{\rm c.m.} <  1.7$ MeV) as compiled in Ref. \cite{Gai13}. We refer the reader to Ref. \cite{Gai13} for a comprehensive review of current data and other data analyses including the most recent analysis of Schuermann {\em et al.} \cite{Sch12}. The current paper relies on the analysis presented in Ref. \cite{Gai13} where it is demonstrated that using $\chi ^2$ analysis the data shown in Fig. 2 aggregate into two distinct groups. We also refer the reader to Ref. \cite{Gai13} for detailed $\chi^2$ analyses that lead us to remove the EUROGAM/GANDI array results \cite{Ham1,Ham2} from the ``world data" and establish the grouping of the data shown in Fig. 2.

\begin{figure}[hbt] 
\begin{center} 
\includegraphics[width=5in]{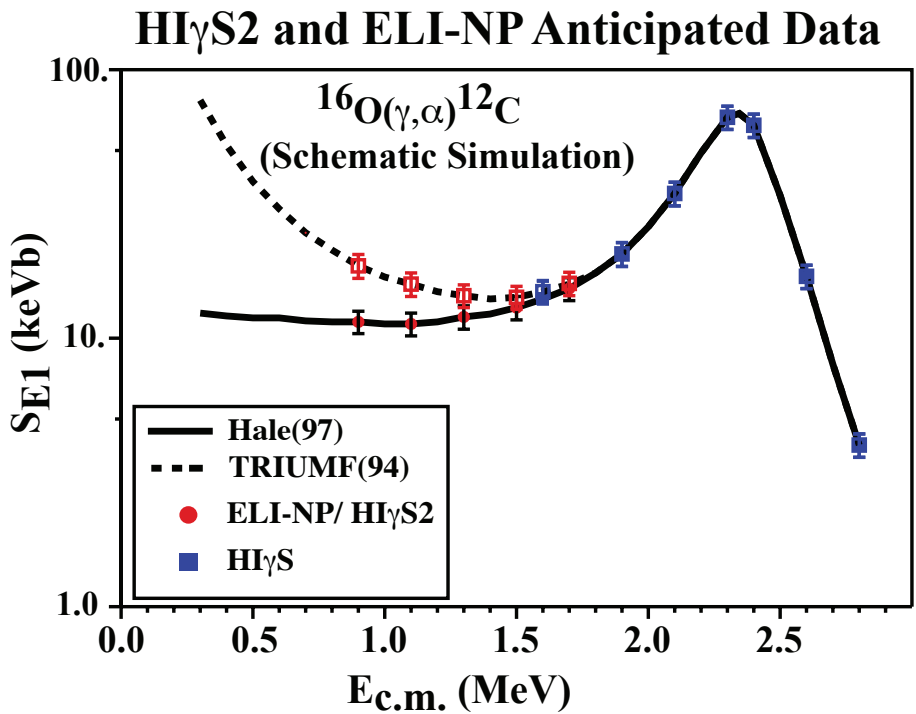} 
\caption{The anticipated (simulated) $S_{E1}$ values that will be measured by the current HI$\gamma$S facility 
(E$_{\rm c.m.} \approx 1.6$ MeV) \cite{Gai} and at the proposed HI$\gamma$S2 and ELI-NP facilities.}
\label{fig:E1}
\end{center} 
\end{figure} 

We conclude that current ``world data" on $S_{E2}$ extracted from angular distributions measured at energies below 1.7 MeV cluster in two distinct groups, leading to two different extrapolations of $S_{E2}(300): \ \approx 60$ or $\approx 154$ keVb. Neither one of these solutions can be favored or ruled out by the current ``world data" of measured angular distributions. In order to resolve this ambiguity in the value of $S_{E2}(300)$ one needs to measure complete and very detailed angular distributions for the $^{12}$C$(\alpha,\gamma)^{16}$C reaction with high accuracy (with binning of 10$^\circ$ or less) at very low energies (below 1.8 MeV). The data at large backward angles are most sensitive to the $E2/E1$ ratio, but such measurements with gamma-ray detectors are challenged by the finite size of the gamma-ray detector and the presence of the beam pipe. 

An experiment is currently in progress at the HI$\gamma$S gamma-ray beam facility to measure the time reverse $^{16}$O$(\gamma,\alpha)^{12}$C reaction using a Time Projection Chamber (TPC) detector operating with $CO_2$ gas \cite{OTPC} and gamma-ray beam \cite{Gai}. Such a measurement with gamma-ray beam and a TPC detector does not suffer from the limitation imposed when using a gamma-ray detector and angular distributions can be measured with any bin size for all angles between 0$^\circ$ and 180$^\circ$ as demonstrated in our recent measurement of the $^{12}C(\gamma,\alpha)^8Be$ reaction \cite{Zim} with an Optical readout TPC (O-TPC) at the HI$\gamma$S facility.

The ambiguity in the value of the extrapolated $S_{E2}(300)$ reported in Ref. \cite{Gai13} resembles the more familiar ambiguity in the value of the extrapolated $S_{E1}(300)$ value where even the data on the $\beta$-decay of $^{16}$N shown in Fig. 18 (and Fig. 16) of \cite{TRIUMF} reveal two minima with identical $\chi^2_\beta$ values at $S_{E1}(300) \ \approx$ 10 keVb and $\approx$ 80 keVb. The small value of the extrapolated $S_{E1}(300) \approx 10$ keVb has been discussed by many authors \cite{Caltech,Redder,Ouellet2,Hale,Gial1} and cannot be resolved by the modern data as shown in Fig. 5 of \cite{Ham1}. In order to resolve this ambiguity in the value of $S_{E1}(300)$ the newly proposed experiments \cite{Gai} must measure complete gamma-ray angular distributions of the $^{12}$C$(\alpha,\gamma)^{16}$O reaction with high accuracy at low energies (around 1.0 MeV). A new measurement of the time reversed $^{16}O(\gamma,\alpha)^{12}$C reaction is in progress using gamma-ray beams from the HI$\gamma$S facility \cite{Gai} and new measurements are anticipated for the proposed new facilities HI$\gamma$S2 at Duke and ELI-NP in Bucharest, Romania. The anticipated sensitivity of this measurement is shown in Fig. 3 using simulated data.

\section{Conclusion}

We conclude that the measured $S_{E2}$ values of the $^{12}C(\alpha,\gamma)^{16}$O reaction bifurcate into two groups extrapolating to $S_{E2}(300) \ \approx$ 60 keVb or $\approx$154 keVb. This newly observed ambiguity in the extrapolated $S_{E2}(300)$ value resembles the familiar ambiguity in the extrapolated $S_{E1}(300)$ value where the small $S_{E1}(300) \ \approx$ 10 keVb solution cannot be ruled out in favor of the large $\approx$ 80 keVb solution. These ambiguities in the extrapolated $S_{E2}(300)$ and $S_{E1}(300)$ values must be considered by practitioners in the field of stellar evolution theory and they must be resolved by the experiments now in progress \cite{Gai}.

\section*{Acknowledgments} 
This work is supported in part by the U.S. Department of Energy, grant number DE-FG02-94ER40870.


\begin{thebibliography}{99}
\bibitem{Gai1} According to Gerry the original temperature scale in the Scientific American article \cite{SciAm} was plotted in keV but it was changed by the request of the publisher to the more familiar scale of Millions of Degrees Kelvin (MK), but it should have been changed to approximately 10 MK leading to the missing factor of 10; e.g. helium burning sets in at approximately 20 keV which is approximately 200 MK.
\end{thebibliography}
\end{document}